\documentclass[]{interact}
\usepackage[utf8]{inputenc}
\usepackage{epstopdf}
\usepackage[caption=false]{subfig}
\DeclareUnicodeCharacter{3B1}{$\alpha$}

\usepackage[natbibapa,nodoi]{apacite}
\usepackage{graphicx}
\theoremstyle{plain}

\theoremstyle{definition}

\theoremstyle{remark}

\begin{document}


\title{Rushing or dragging? An analysis of the ``universality'' of correlated fluctuations in hi-hat timing and dynamics}

\author{
\name{Oliver Gordon, Dominic Coy, Jack Matthews, Easel Kandola-McNicholas, Owain
	Llewellyn, Adeel Bokhari, and Philip Moriarty\textsuperscript{*}\thanks{\textsuperscript{*}CONTACT Philip Moriarty. Email: philip.moriarty@nottingham.ac.uk}}
\affil{School of Physics \& Astronomy, University of Nottingham, Nottingham NG7 2RD, UK}
}

\maketitle

\begin{abstract}
A previous analysis of fluctuations in a virtuoso (Jeff Porcaro) drum performance [R\"as\"anen et al., PLoS ONE 10(6): e0127902 (2015)] demonstrated that the rhythmic signal comprised both long range correlations and short range anti-correlations, with a characteristic timescale distinguishing the two regimes. We have extended Rasanen et al.'s approach to a much larger number of drum samples (N=132, provided by a total of 58 participants) and to a different performance (viz., Rush's \textit{Tom Sawyer}). A key focus of our study was to test whether the fluctuation dynamics discovered by R\"as\"anen et al. are ``universal'' in the following sense: is the crossover from short-range to long-range correlated fluctuations a general phenomenon or is it restricted to particular drum patterns and/or specific drummers? We find no compelling evidence to suggest that the short-range to long-range correlation crossover that is characteristic of Porcaro's performance is a common feature of temporal fluctuations in drum patterns. Moreover, level of experience and/or playing technique surprisingly do not play a role in influencing a short-range to long-range correlation cross-over. Our study also highlights that a great deal of caution needs to be taken when using the detrended fluctuation analysis technique, particularly with regard to anti-correlated signals.
\end{abstract}


\section{Introduction}

It was established more than forty years ago that the power spectra of fluctuations in musical pitch and loudness are of a 1/\textit{f} character (where \textit{f} represents frequency) \citep{Vos1975}. This was a fascinating and exciting discovery, given that power law scaling of the 1/\textit{f} type is ubiquitous in nature, spanning a remarkably wide variety of organic and inorganic systems (and their associated length- and time-scales). From flicker noise in electronic and ionic devices \citep{Balandin2013,Zorkot2016} to fluctuations in heartbeats \citep{Ivanov2001} and brain activity \citep{Bedar2006}, 1/\textit{f} scaling is indicative of systems that could be described as being in a ``Goldilocks'' regime in terms of their temporal correlations: not entirely uncorrelated (i.e. white noise, yielding a flat power spectrum) but lacking the strong correlated ``memory'' of Brownian noise \citep{Mao2002}. Instead, systems exhibiting 1/\textit{f} noise have correlations which fall between these two extremes. 

In an engaging article for his celebrated Mathematical
Games column in \textit{Scientific American} \citep{Gardiner1978}, Martin Gardiner explained back in 1978 that music that falls between the white and brown(ian) noise limits, i.e. whose fluctuations are of 1/\textit{f} character, appeals due to its moderately correlated (``just right'') variations. Jazz, rock, and classical music were each found to follow 1/\textit{f}-type scaling below 1 Hz, i.e. on long time scales, in terms of fluctuations in both loudness and pitch in Voss and Clark's analysis. Subsequent studies argued that subtleties in the scaling of the fluctuations could be used to distinguish one genre of music from another. In particular, it has been claimed \citep{Biegerelle1999} that the fractal dimension of time series samples of different 
forms of music - instead of the power law character of the associated power spectra - could be used to distinguish, for example, heavy metal from jazz or a string quartet.
(It is perhaps somewhat over-optimistic, however, to base this distinction on values of fractal dimension quoted to five or six 'significant' places, as in \citep{Biegerelle1999}). The relationship between the fractal dimension of a time series and the 1/f$^{\beta}$ character of its associated power spectrum is discussed in Methods below. For now, note that white noise is associated with a $\beta$ value of 0 (in other words, the power spectrum is independent of frequency), whereas brown noise is characterised by a $\beta$ value of $\sim$ 2.

If fluctuations in pitch and loudness are correlated, one might quite reasonably ask if rhythm and timing are subject to similar correlations. Hennig and co-workers focused	on exactly this question, analysing variations in inter-beat intervals for a variety of drum patterns played both with and without a metronome (or click track)\citep{Hennig2011,Hennig2012}. As	they point out, there are a number of long-standing questions on the neuronal	underpinnings of human response times on millisecond time scales; studies of drumming can play a key role in elucidating the associated dynamics. Of especial interest in the context of this paper was Hennig \textit{et al}.'s discovery of long range correlations in interbeat intervals \citep{Hennig2011}; small fluctuations in timing at a particular point in time have a significant influence on a drummer's ``output'' for as long as tens of seconds after the event.

The work we describe here was, however, largely inspired by R\"as\"anen \textit{et al}.'s recent analysis of an iconic and challenging drum pattern: Jeff Porcaro’s single-handed 16$^{th}$ note hi-hat groove on Michael McDonald’s 1980s classic, \textit{I Keep Forgettin}’ \citep{Ras2015}. A
cross-over from short range anti-correlations to long-range correlated interbeat fluctuations was found to be a key characteristic of Porcaro's performance on that track. As the authors discussed, this has fascinating implications for both the biophysics and psychology of our appreciation of ``groove'', and the potential mimicking of similar correlations in attempts to “humanise” computer-simulated drum tracks.

A key motivation for our study was to elucidate the extent to which the crossover of correlation regimes discovered by R\"as\"anen \textit{et al}. in Porcaro's playing is a feature of drumming in general. A variety of fascinating questions spring to mind. Do all drummers produce similar trends in fluctuations? Or is the crossover characteristic of	only a particular style or level of experience? To what extent are the correlations in	fluctuations indicative of a particular drummer? And, ultimately, could a drummer's style be “dialled in” to a drum machine by simulating the correct trends in the fluctuation dynamics?

\section{Methods}
A methodology quite similar to that introduced by R\"as\"anen \textit{et al.} but with key modifications to the data acquisition and processing protocols (for reasons discussed below) was adopted. Data were acquired through two primary methods:(i) crowd-sourcing via social media (YouTube, Twitter, Facebook) and, (ii) personal contacts with musicians involved in music societies, orchestras, and bands. These two types of source produced a very significant amount of data which would have been exceptionally difficult to analyse “by hand” in the manner adopted by R\"as\"anen \textit{et al.} in order to isolate the hi-hat beats in Porcaro's track. We instead developed an automated process that located the beats in .wav sample files of the various drum patterns. As compared to R\"as\"anen \textit{et al}.'s analysis, we had the distinct advantage that isolation of the hi-hat pattern was substantially more straight-forward. The vast majority of our analyses were based on tracks for which only the hi-hat was played along with the track of interest; ``extraneous'' signals from instruments other than the drums were eliminated in a way that is impossible for the \textit{I Keep Forgettin'} recording.	

Rush's \textit{Tom Sawyer} was selected for analysis in our case due to it being another challenging single-handed sixteenth note hi-hat pattern played at a similar tempo. It is also  what is best described as a virtuoso performance. Neil Peart, Rush's drummer, has stated that \textit{Tom Sawyer} is the song he finds most challenging to play live. In hindsight, our choice of a challenging song such as \textit{Tom Sawyer} was perhaps somewhat ill-advised in the context of producing well-defined correlations over extended periods of time. A more straight-forward beat may better elucidate the extent of correlated patterns in drumming, where factors such as stamina are less of an issue. (The 88 bpm single-handed sixteenth note pattern in \textit{Tom Sawyer} can be difficult to sustain for even experienced drummers). We return to this point in the conclusions.

\subsection{Sourcing drum tracks}	
Initially, data were sourced via a video uploaded to the Sixty Symbols YouTube channel \citep{60s2015}. Although 78 different tracks were submitted in response to the Sixty Symbols video, the majority of these were unfortunately not of a sufficient duration to enable a robust investigation of the crossover to the long range correlation regime. Those tracks nonetheless provided valuable information regarding the reliable extraction of interbeat timings and amplitudes.

In addition to the track duration problem, crowd-sourcing data recorded under widely varying conditions (and with substantial variations in audio quality) represented a major challenge with regard to ensuring consistency of analysis and processing. We therefore decided to switch to a protocol whereby drummers were recorded under very similar conditions (see following section). In this case, 22 drummers were identified via personal communications with musicians and music societies, orchestras, and bands. All drummers who contributed are listed in the Acknowledgements section at the end of the paper.	

In order to provide both control data and an analysis of the extent to which drumming experience might play a role in determining correlations in interbeat fluctuations, a total of 22 high school students, none of whom had drumming experience, were asked to play along with a metronome running at 88 beats per minute, i.e. the tempo of \textit{Tom Sawyer}.

\subsection{Recording and pre-processing}
In the \textit{Tom Sawyer} drum pattern of interest here, each bar consists of four beats. These four beats represent four crotchets, or 16 semiquavers (i.e. sixteenth notes). To record samples, twenty-two musicians were asked to play semiquavers for a minimum of two minutes on a closed hi-hat whilst simultaneously listening to the original recording of the song (from the album \textit{Moving Pictures}). Each performance of semiquavers was then repeated once with a double-handed method, once with a single-handed method, and once with a technique of the drummer's choice but in as metronomic and unaccented a fashion as possible. This was all performed with the drummer's own sticks and, other than the metronomic recording, they were asked to play as if they were giving a live	performance, instead of just trying to keep time. Our aim was to facilitate the drummers' playing in a style that they prefer in order to better reflect a real-world	performance. Nonetheless, the drumming in our study was of course carried out in an artificial environment which is very different from that of a concert performance, a point to which we also return in the conclusions. These data were complemented by hi-hat-only tracks received via the Sixty Symbols crowd-sourcing route.

For the sample of non-drummers, only the metronomic option was requested. Many found it difficult to sustain the sixteenth note pattern for any time using a single-handed technique and either switched to a double-handed technique or resorted to playing	eighth, rather than sixteenth, notes. Finally, a small number of tracks were recorded with Aerodrums \citep{Aerodrums} rather than physical drums. In this case there is no physical feedback from the surface of the hi-hat (because there is no physical drum kit). Instead, the Aerodrums system exploits light-tracking to monitor the positions of the tips of the drum sticks and trigger the appropriate samples accordingly. One of the authors (PM) and one of the inventors of the Aerodrums system (Richard Lee) recorded both the \textit{Tom Sawyer} and the \textit{I Keep Forgettin’} pattern on Aerodrums (see Supplementary Information). The biophysical	response of the drummer is clearly very different for a physical kit vs Aerodrums\footnote{It is worth noting that although the lack of physical bounce from a solid drum surface might at first appear to limit the application of Aerodrums (due to the system not fully mimicking a ``real world'' kit), with the correct technique, drum bounce can be simulated very accurately by allowing the (loosely held) stick to rebound off the palm of the hand. Indeed, many drum teachers recommend this approach even when playing on a physical kit as it eliminates any reliance on the stiffness and response of the physical surface.}. These aspects of the Aerodrums-related recordings are distinct from the overall scope and	objectives of the research and we are certainly not in a position, given the extremely limited sample size (two Aerodrummers) to draw any type of conclusion on biophysical response. We will instead return to a comprehensive analysis of the dynamics and correlations of Aerodrums vs physical drums in a later paper.	

For the core analysis on which the majority of this paper is based, the twenty-two drummers were recorded under identical conditions. Immediately before the track started, eight crotchets were played to each musician via a click track. After hearing	four of the crotchets, each musician then played another four crotchets – one on each beat. This protocol ensured that each musician began in time with the music and did not have to catch up or slow down to the beat at the start of the track (which would have significantly skewed the results). It also provided a very distinct starting location for the beat onset detection algorithm discussed below. To avoid false onset detection, recordings were carried out in a quiet room free of significant background noise and reverberation, with the microphone facing the hi-hats. The microphone was also adjusted to avoid saturation (which would have made the accurate detection of onsets problematic). Each drum track was recorded in stereo at a sampling frequency of 44.1kHz, converted to mono, and then saved in an uncompressed 32 bit .wav file format.	

\subsection{Beat breakdown: Onset detection}
We used a simple time domain treatment of the recorded drum signals to detect interbeat timings and amplitudes. A hard low pass filter (fifth order Butterworth; cut-off between 50 Hz and 200 Hz) was
used to smooth the signal and provide an envelope for the hi-hat pulses. Matlab's standard peak detection algorithm (\textit{findpeaks()}) was then sufficient to detect the
hi-hat impulses (see Fig.1), subject to the following restrictions. This differs somewhat from the Fourier space approach adopted by R\"as\"anen \textit{et al.} in that we found that generating a spectral analysis of the signal did not produce more reliable onset	detection. (Unlike R\"as\"anen and co-workers, however, we were generally in the fortunate
position of having a hi-hat only track so that frequency spectrum “crowding” due to other instruments was not an issue). In addition, we found that the use of signal maxima, rather than minima, to define onsets led to fewer false positives.	

A windowing-based method was used to limit the beat locations to a specific temporal range. To find the $i + 1th$ onset, the expected time between onsets, $t_{exp}$, was first calculated from the tempo of the piece. The version of Tom Sawyer used in our study is played at 88 bpm\footnote{In other words, there are 88 quarter notes (crotchets) per minute.} and thus $t_{exp}$ is 0.170 s. The algorithm then added $t_{exp}$ to $t_i$ so as to shift the time “stamp” to the next position. A peak search algorithm hunted within a time window which is $\pm$ 34 ms wide about the value of $t_i + t_{exp}$. (The window width was chosen on the basis of R\"as\"anen \textit{et al}.’s hi-hat detection work.) The peak in this window was found and its corresponding time recorded as $t_{i+1}$.	

The time intervals between onsets, $\tau_i$, are then easily calculated:	
\begin{equation}
\tau_i= t_{i+1} - t_i
\end{equation}
	
Erroneous time intervals of $\tau >$ 34 milliseconds were then removed, as these time differences correspond to missed onsets. (These missing onsets have a number of origins ranging from a drummer inadvertently skipping a beat to an extended open hi-hat
note). This left $N$ detected onset intervals. The number of intervals expected was calculated by dividing the total time in which sixteenth notes are played by $t_{exp}$. This value was then used with $N$ to calculate the percentage of onsets detected.

\begin{figure}
	\centering
     \includegraphics[scale=2.0]{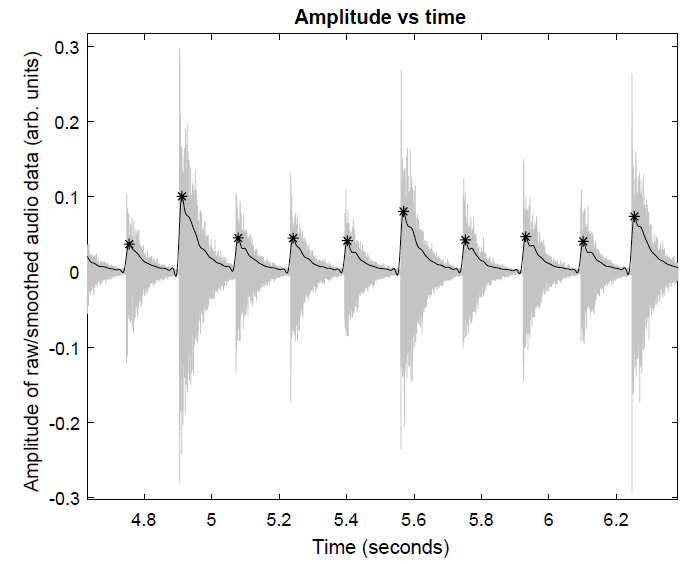}
	\caption{Illustration of peak detection algorithm. Original drum sample shown in gray; low-pass filtered (200 Hz cut-off; fifth order) data shown as black line. Identified peaks are highlighted with a * symbol.} \label{PeakDetect}
\end{figure}

The positions of those hi-hat peaks were subsequently used with the original, unfiltered, data to extract the amplitude of the hi-hat signal at that time. Fig.\ref{PeakDetect} shows an example of the output of the peak detection code.

There are two obvious objections that could be raised to this peak-finding strategy: (i) to what extent does any phase shift introduced by the filter lead to artefacts and errors in determining the times at which beats occur (particularly given that the signal is heavily filtered to minimise false peak detection), and (ii) does the heavy modification
of the peak shape (due to the removal of higher Fourier components via the filter) similarly produce timing artefacts?	 In order to determine the efficacy and robustness of the peak detection code we therefore tested it thoroughly with simulated drum fluctuation samples of various ``colours''. We show examples of the analysis of pink and blue noise drum samples in the following section. Other examples (white, brown, and uncorrelated noise of various types) are included in the supplementary material, along with the simulated drum samples themselves (as wav files).	

\subsection{Detrended fluctuation analysis: the Hurst exponent and fractals}

Almost seventy years ago, Hurst introduced what came to be an influential method of quantifying correlations in time series dominated by stochasticity: the rescaled range heuristic \citep{Hurst1951}. Hurst's approach involves splitting a time series into a set of adjacent windows. The range of the fluctuations within each of those windows is determined (and scaled to the standard deviation) as a function of the size of the window.	The rescaled range (i.e. the range divided by the standard deviation) that gives the method its name is related to the window size via a power law. The exponent for this power law (the Hurst exponent) is denoted $\alpha$. The fractal dimension of a time series, $D$, and its Hurst exponent, $\alpha$, have the following relationship: $D = 2 - \alpha$.

As discussed in the Introduction, there is another exponent, $\beta$, which describes the frequency dependence of the power spectrum of the signal, $1/f^{\beta}$. There is, in turn, a simple relationship between $\alpha$ and $\beta$, namely $\beta$ = 2$\alpha$ - 1. Fo example, an uncorrelated (white noise) signal which has a flat power spectrum independent of frequency is associated with $\beta$ = 0. Thus, $\alpha$ = 0.5 for white noise. A value of $\alpha$ in the range
$0.5 < \alpha < 1.5$ is instead indicative of long-range correlations, while $0 < \alpha < 0.5$ is characteristic of anti-correlations. It turns out that many natural systems are associated with a Hurst exponent of 0.7, which is often described as correlated noise.

Detrended fluctuation analysis (DFA) was designed \citep{Peng} to improve on Hurst's rescaled range method by removing trends in each of the signal windows. (A thorough review and in-depth critique of the development and application of DFA is given in \citep{Bryce}.) The DFA technique expands on Hurst's approach by fitting the data to a line or a curve (usually a polynomial) within the window of interest and subtracting this - hence ``detrending'' the data - before the fluctuations are analysed. Following R\"as\"anen \textit{et al.}'s approach, we used DFA to analyse both interbeat and amplitude variations in sixteenth note hi-hat patterns. In the following we outline our approach to the application of DFA to scaling fluctuations in interbeat timings. (The analysis of amplitude fluctuations was carried out in the same manner).

We first calculate the cumulative deviation from the mean, $\delta_i$, as follows:
\begin{equation}
\delta_i=\sum_{j=1}^{i} (\tau_j - \bar{\tau})
\end{equation}

\noindent where $\bar{\tau}$ is the mean value of the interbeat intervals. The summation in Eqn. 2 is repeated for $1 < i < N$, where $N$ is the total number of intervals. A small modification
of Eqn.2 allows us to calculate the temporal drift, $d_i$, from the mean:
\begin{equation}
d_i=\sum_{j=1}^{i} (\tau_j - i\bar{\tau})
\end{equation}

\begin{figure}[b!]
	\centering
	\includegraphics[scale=1.5]{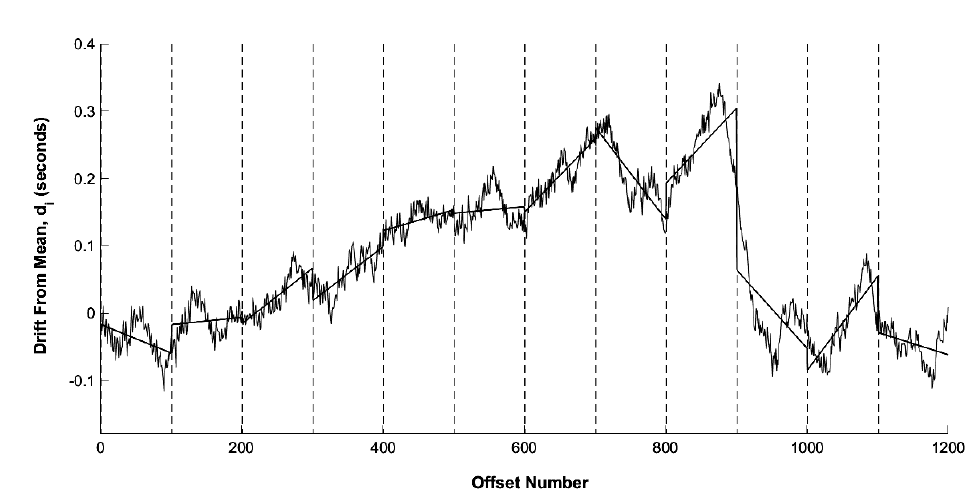}
	\caption{An example of linear detrended fluctuation analysis. The cumulative sum of onset differences from the mean interbeat interval (i.e. Equation 2) is plotted as the solid line. An integer number of windows of width $s$ are then created. Here $s=100$ and the edges of the windows are shown as vertical dashed lines. A linear fit to the drift data, $\delta_{FIT}$ in Equation 4, is then made within each window. This is repeated for windows of size $s$ = 5 to (typically) $s$ = 160 to produce ln $F$($s$) vs ln $s$ graphs of the type shown in the Results section.} \label{DFA}
\end{figure}

Next, the $N$ values of $\tau$ are split up into windows of equal size, $s$, resulting in a total of $N/s$ windows. If there are insufficient $\tau_i$ points to form a window, the excess points were discarded so to ensure an integer number of windows. A line was then fitted to the interbeat values in each window and the data points for the fit, $\delta_{FIT_i}$, recorded. The process is repeated for window sizes typically in the range $5 < s < 160$. (Depending on the duration of the sample, however, the upper limit on occasion was reduced.) An
example of the data treatment for $s = 100$ is shown in Fig.\ref{DFA}.

The root-mean-square (RMS) fluctuations within each window are then calculated as follows:
\begin{equation}
F_s(k)=\sqrt{\frac{1}{s}\sum_{i=ks+1}^{ks+s}\big{[}(\delta-\delta_{FIT})_i\big{]}^2}
\end{equation}

\noindent where $k$ is an integer. $F_k(s)$ is calculated for $0 < k < s - 1$ and, typically, $5 < s < 160$. This
produces $N/s$ values of $F_k(s)$ for each value of $k$. A mean is taken of this set of $s$ values to yield $F(s)$, i.e. the mean of the mean square fluctuations for a window size, $s$. $F(s)$ is related to the Hurst exponent, $\alpha$ as follows: $F(s) \propto s^{\alpha}$. The gradient of a graph
of ln $F(s)$ vs ln $s$ therefore yields the value of $\alpha$.

\section{Results and Discussion}

\begin{figure}
	\centering
	\includegraphics[scale=2.0]{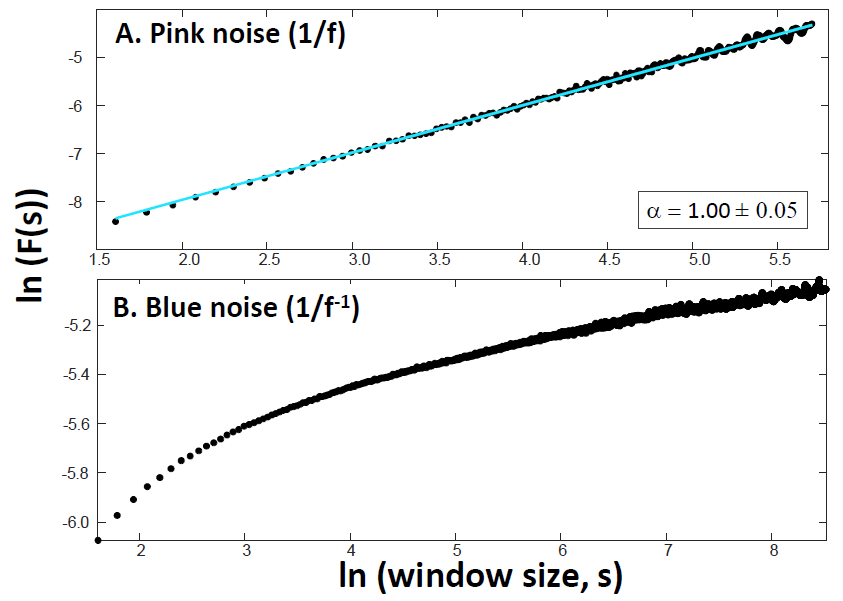}
	\caption{Detrended fluctuation analysis (DFA) of simulated hi-hat samples where the fluctuations (in both interbeat timing and amplitude) followed (a) $1/f$, i.e. pink noise, scaling, or (b) $1/f^{-1}$, i.e. blue noise, scaling. (Only the interbeat timing fluctuations are shown here.) The DFA treatment for $1/f$ fluctuation scaling not only produces a single straight line (as expected on the basis of the power law scaling used to simulate the hi-hat sample), it also returns the correct value for the exponent, within an error bar estimated solely from the scatter in the points comprising the log-log plot (and calculated within Matlab’s polyfit framework.) DFA analysis of the hi-hat sample encoded with blue noise fluctuations, however, exhibits a great deal of curvature. Graph (b) is the log($F(s)$) vs log ($s$) plot for 32768 consecutive hi-hat beats. Only for very large window sizes does the value of $\alpha$ start to converge to a value reasonably close to zero: 0.059 $\pm$ 0.009} \label{noise}
\end{figure}

To first test the reliability of the peak detection and DFA algorithms described in the previous section, we simulated hi-hat samples whose fluctuations were described by one of a range of noise colours. Audio files of hi-hat beats where the interbeat timings and/or amplitudes fluctuated according to a $1/f^{\beta}$ power law ($ -1 < \beta \leq 2$) were generated using Matlab's digital signal processing toolbox to produce noise of the appropriate colour. For pink noise (i.e. $\beta = 1$), Fig.\ref{noise}(a), it is clear that DFA can be used to accurately extract the appropriate power law dependence from the data. We similarly found that DFA performed well for values of $\beta$ in the range $0 \leq \beta \leq 2$ (See Supplementary Material.)	

In contrast, DFA performed rather poorly on data of strongly anti-correlated character. (We note at this point that DFA is limited to the detection of values of $\alpha$ that fall within the range $0 < \alpha < m + 1$, where $m$ is the order of the polynomial fit used in the DFA process \citep{Kiyono2015}. In our case, and in common with many other published applications of DFA, we use a linear fit. DFA employing linear regression to remove trends is thus restricted to the analysis of data for which $0 < \alpha < 2$.) The data in
Fig.\ref{noise}(b) are for hi-hat fluctuations of blue noise character, i.e. the ``opposite'' of pink noise, where the power spectrum scales directly (linearly), not inversely, with frequency. In this case $\beta = -1$, and thus $\alpha$ is 0. Even in the limit of an exceptionally large number of samples - which was 32768 in this case, representing a total duration of a little over 90 minutes - DFA not only fails to extract the correct value of $\alpha$ but the data are highly non-linear.

Although the true value of $\alpha$ for Fig.\ref{noise}(b), i.e. 0, strictly lies outside the valid range for linearly detrended DFA, and is thus a worst case example, the lack of linearity we observe in that figure is also an issue for values of $\alpha$ somewhat greater than 0 (see Supplementary Material for other values of $\alpha < 0.25$; R\"as\"anen and co-workers have similarly recently found issues with the reliability of DFA for values of $\alpha$ towards the limits of the valid range of applicability of the technique (\cite{Esa}). The lack of linearity in  Fig.3(b) is clearly a troublesome deficiency of the DFA method when it comes to	treating anti-correlations. This issue would seem to be compounded by the much greater sensitivity of anti-correlations (as compared to correlations) to data removal \citep{Ma2010} prior to the application of DFA. In the event that the hi-hat drum track (or appropriate
other drum track, e.g. ride cymbal or bass drum(s)) cannot be isolated, there will clearly be missing data when the hi-hat “pulse” is lost within the musical backdrop	created by the other instruments (including drums other than that of interest).

Other key deficiencies in DFA have been highlighted by a number of authors including \citep{Bryce}, who argue that DFA is simply not ``fit for purpose'' in many cases as it can both inadvertently introduce bias and, despite being designed specifically to address the issue, does not protect against non-stationary signals and trends. A decade earlier, Hu \textit{et al.} \citep{Hu} had also flagged up issues with DFA's handling of trends, showing that crossovers can result in the scaling behaviour of the noise and the apparent scaling of the trend(s) in question. In a similar vein to Bryce and Sprague, although arguably more damning, \citep{Dingwell} made the point that DFA is prone to “false positive” results, citing previous work \citep{Maraun} (among others) who heavily criticised the fundamental “tenet” of DFA, arguing that ``scaling cannot be concluded from a straight line fit to the fluctuation function in a log-log representation'' (their words.) 

Nonetheless, DFA remains a very popular technique for the extraction (spurious or otherwise) of scaling exponents from noise and fluctuations. With the critiques and reservations listed above in mind, we now turn to a discussion of the application of DFA to the \textit{Tom Sawyer} drum data we acquired. In total, we analysed in detail 132 drum samples from 36 participants who had drumming experience\footnote{Of these 36 drummers, 22 were recorded under identical circumstances as described in the Methods section, while 14 submitted their tracks via the Sixty Symbols “crowd-sourcing” strategy also discussed above.}, and from 22 high school students who had not drummed previously. The total of 132 samples also includes single-handed and double-handed versions of the 88 bpm sixteenth note pattern, and spans hi-hat tracks recorded while listening to the version of the track recorded by Rush for their \textit{Moving Pictures} album and to a metronome, respectively.

\begin{figure}
	\centering
	\includegraphics[scale=2.5]{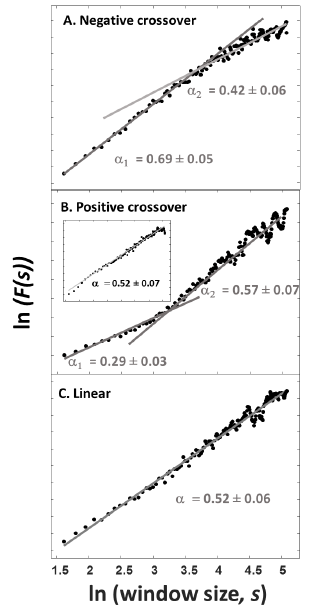}
	\caption{Examples of the three key classes of interbeat fluctuations found via detrended fluctuation analysis. In each case we plot the natural logarithm of $F(s)$ vs the natural logarithm of $s$, the window size. \textbf{(a)} ``Negative crossover'', where data representing a higher value of cross over to a lower value (see text for a discussion in the context of artefacts arising from the DFA technique); \textbf{(b)} ``Positive crossover'' of the type found by R\"as\"anen \textit{et al.} The inset shows the ln $F(s)$ vs ln $s$ plot for the amplitudes of the beats. In this case there is no correlation in amplitude – the value of is 0.52 $\pm$ 0.07, representing white noise. (Note the ``transient'' for small values of $s$. This is a common feature of DFA analysis of white noise fluctuations.); \textbf{(c)} entirely linear behaviour, with a value of $\alpha$ in this case that is again indicative of a lack of correlation in interbeat timings. See Table 1 for the relative frequency of occurence of each of these types of curve. Linear behaviour is by far the most common.} \label{Fs}
\end{figure}

Fig.\ref{Fs} shows a representative subset of $F(s)$ plots produced via the data treatment described in the previous section. The plots fall into three classes which we describe as linear, negative crossover, and positive crossover. We use the term ``positive crossover'' to denote an $F(s)$ plot that behaves in the manner described in \citep{Ras2015}, i.e. where the data for lower window widths are associated with a smaller value of $\alpha$ than those for larger values of $s$. In other words, there is a cross-over in the behaviour of the
fluctuations on short and long timescales.

As an additional test of the efficacy of our peak detection algorithm, we also translated the interbeat timings and amplitudes very helpfully provided by R\"as\"anen \textit{et al.} for their analysis of Porcaro's performance on \textit{I Keep Forgettin’} into a hi-hat track (i.e. a .wav file) which we subsequently analysed to ensure that we recovered the crossover in exponents. This analysis is covered in the supplementary material. We note here, however, that there is quite some flexibility in choosing the precise location of the crossover point in the log-log plot. \citep{Ge}, among others, have stressed the lack of an objective heuristic for locating the cross-over point beyond “eye-balling” the data\footnote{In an attempt to make this approach more systematic, OG and DC developed an algorithm based around an analysis of the variation in gradient at the extremes of the window size axis. Unfortunately, this had limitations with regard to just how to define those extremes.}. This subjectivity in choice of cross-over contributes another uncertainty to the determination of the values of $\alpha$,
which, in our experience, can often be comparable to the uncertainty due to the scatter in the points. More broadly (and more worryingly), the DFA algorithm itself has been known to introduce entirely spurious crossovers in scaling \citep{Ignaccolo}.	

Only for a very small number of drum samples (4 out of the total of 132 audio files, i.e. 3 \% of the data) did we see a crossover in the value of $\alpha$ on shorter vs longer time-scales that was consistent with the behaviour observed in \citep{Ras2015}, having what we describe as positive crossover. In the particular case of Fig.4(b), the value of $\alpha$ switches from 0.29 ($\pm$ 0.05) to 0.57 ($\pm$ 0.07). In identifying credible instances of scaling crossover behaviour we rejected data where, due to an audio file which was not of sufficient duration, spurious oscillations appeared in the ln $F(s)$ vs ln $s$ plot for large window sizes. An example of this type of oscillatory artefact is provided in the Supplementary Material.

Table 1 below shows the breakdown of the data into the three categories of scaling shown in Fig.4: linear, positive crossover, and negative crossover. We find no correlation between drumming experience and either the value of the scaling exponent or the	presence/lack of crossover behaviour in the associated ln $F(s)$ vs ln $s$ plot. This is entirely in line with the findings of \citep{Hennig2011}, who similarly found that it was not possible to characterise a given musician using a value of $\alpha$ or $\beta$; an individual performed differently on repeated attempts at a given task.

\begin{table}[b!]
\begin{center}
\begin{tabular}{|l|c|c|c|}
	\hline 
	& \textbf{Positive Crossover} & \textbf{Negative crossover} & \textbf{Linear} \\ 
	\hline 
	\textbf{Interbeat timings} & 3\% & 12\% & 85\% \\ 
	\hline 
	\textbf{Amplitudes} & 2\% & 0\% & 98\% \\ 
	\hline 
	\hline
\end{tabular}
\end{center}
	\caption{Categories of scaling for all 132 drum samples we have analysed. “Positive crossover”, “negative crossover”, and “linear” are as defined in Fig.\ref{Fs}. As described in the text, we find no robust correlation between the crossover/lack of crossover behaviour and drumming experience.} 
\end{table}

\begin{figure}
	\centering
	\includegraphics[scale=1.75]{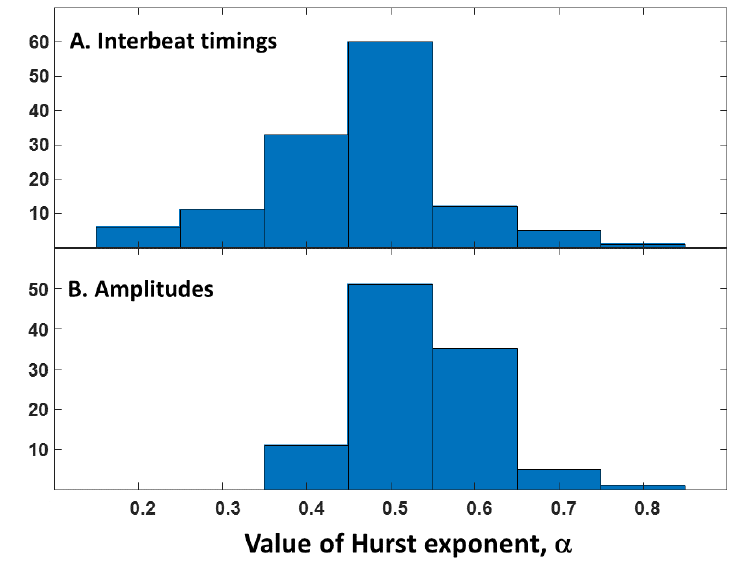}
	\caption{Histograms of the Hurst exponents for fluctuations in \textbf{(a)} interbeat timings, and \textbf{(b)} amplitudes extracted from detrended fluctuation analysis of drum tracks recorded by drummers either playing along to \textit{Tom Sawyer} or to a metronome running at 88 bpm. The histograms include exponents extracted from samples for both single-handed and double-handed styles of playing the sixteenth notes. We found no difference in exponent for the different playing styles. Note that these histograms do not include those participants who had no drumming experience.} \label{exponents}
\end{figure}

There were many drummers who did not produce any crossover in the scaling of the fluctuations - the ln $F(s)$ vs ln $s$ plot could be more than adequately fit with a single	straight line. Moreover, the value of $\alpha$ for these cases was often indistinguishable from 0.5 within experimental error, indicating a lack of correlation on any time scale. In principle, this white noise characteristic might perhaps be expected for drummers with little finesse or ``groove'' in their playing and therefore could (again, in principle) depend on years of experience and/or ``track record'' of drumming. However, while there was certainly variation in the exponents we determined (see Fig.\ref{exponents}), we found no robust link between years of drumming experience and the value(s) of α derived from detrended fluctuation analysis of the hi-hat recordings.

Figs.\ref{exponents}(a) and (b) are histograms for interbeat timings and amplitudes, respectively, for all drummers (i.e. we have excluded all participants who did not have prior drumming experience). As noted in the figure, the error bar for a given value of $\alpha$, as derived from the scatter in the points of the ln $F(s)$ vs ln $s$ plot, is in the 10\% - 20\%
range. In other words, at the upper end of that range the error bar could span not just one but two bins of the histogram. The spread in the values of the Hurst exponent observed for the interbeat timings is somewhat wider than that derived from the
fluctuations in amplitudes.

As is clear from Fig.\ref{exponents}, the range of Hurst exponents for both interbeat timings and amplitudes is centred on a value of 0.5, i.e. a value indicative of a lack of correlation on any time scale. (Note that Fig.\ref{exponents} does not distinguish between values of $\alpha$ associated
with a simple linear ln $F(s)$ vs ln $s$ plot, as compared to the crossover behaviour represented in Figs.\ref{Fs}(a) and (b); we have included all values of $\alpha$ in the histograms.)
Although, as outlined above, our findings are in line with those of \citep{Hennig2011} in terms of the lack of a characteristic $\alpha$ value for a given drummer, we see distinct differences with the results of that work when it comes to the mean and range of the exponents extracted from detrended fluctuation analysis of the data.

\begin{figure}
	\centering
	\includegraphics[scale=1.5]{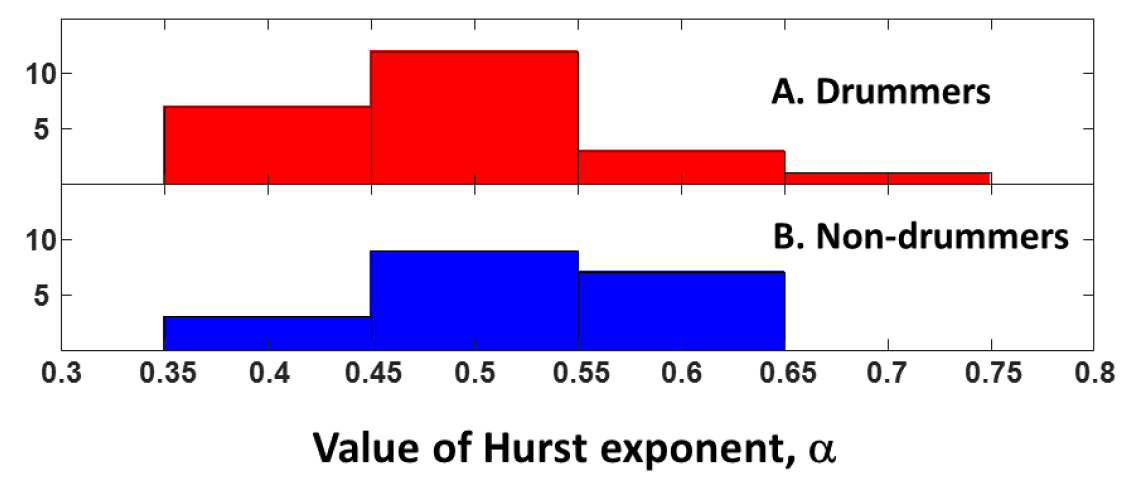}
	\caption{Histograms of the Hurst exponents extracted from drum tracks recorded while playing along to a metronome (88 bpm) for participants with \textbf{(a) }drumming experience, and \textbf{(b)} no drumming experience.} \label{experience}
\end{figure}

In \citep{Hennig2011}, the values of $\alpha$ for both complex and periodic patterns (or for tapping with a stick) fell in the range 0.8 - 1.1 (albeit for a sample size of three drummers). As is clear from Fig.\ref{exponents}, we find no evidence at all for strong long range correlations of pink noise character (i.e. $\alpha \approx 1$) in any of the drum samples we analysed. The reasons for this discrepancy are of course more than likely related to the precise conditions under which the drumming experiments were carried out in the respective studies. In our case, the drummers (and non-drummers) each played only the hi-hat while listening to the recording of \textit{Tom Sawyer} or a metronome running at 88 bpm; the only drum that was being played was the hi-hat. This is likely to have key implications with regard to ``groove''. Moreover, the drummers were asked to maintain a sixteenth note pattern throughout all of \textit{Tom Sawyer} (or for as long as they could drum along to the track.) This includes both the 4/4 and 7/8 sections of the song. This change in time signature could conceivably “scramble” long range correlations. Nonetheless, even for drumming to a metronome we surprisingly found no difference (outside the large error bars) between experienced drummers and those with no prior drumming experience (Fig. \ref{experience}).

\section{Conclusions and Future Work}

Under the conditions under which the drum tracks that we analysed were recorded, we find that only 3\% of the recordings exhibit a positive crossover in scaling exponent of the type associated with Jeff Porcaro's drumming on \textit{I Keep Forgettin'}, as described in \citep{Ras2015} Moreover, and more significantly, we find no evidence for robust long range correlations, i.e. where the value of the Hurst exponent is $\sim$ 1. Instead, across all 132 drum tracks we have analysed (using an automated version of the analysis strategy put forward by \citep{Ras2015}), the vast majority of drummers produce	uncorrelated fluctuations (i.e. $\alpha = 0.5 \pm 0.1$) in both interbeat timings and hi-hat amplitudes. While there are exceptions, with both anti-correlated ($\alpha < 0.5$) and long-range correlated ($\alpha > 0.5$) behaviour observed on occasion (and, even more occasionally, together), we find no link between drumming experience and the presence of correlated or anti-correlated fluctuations. In other words, the Hurst exponent is performance-, rather than performer-, dependent, in good agreement with previous observations (for smaller sample sizes \citep{Hennig2011, Hennig2012}.

In common with a considerable number of other studies,  including \citep{Bryce, Dingwell, Maraun, Ge, Ignaccolo}, we encountered deficiencies and artefacts in detrended fluctuation analysis. In our case, simulated drum tracks having strongly anti-correlated fluctuations in interbeat timing and/or hi-hat amplitudes (i.e for value of $\alpha$ less than $\sim$ 0.25) produce spurious crossover (i.e. non-linear) behaviour in log-log plots of $F(s)$ vs $s$. This raises significant concerns regarding the analysis of anti-correlated fluctuations in drumming via DFA. DFA performed rather better on tests with simulated drum tracks having correlated fluctuations.	

Given that our study supports previous work showing that the scaling of fluctuations in drumming is critically dependent on the details of a given performance rather than being characteristic of a specific performer, an obvious future study, complementing the analysis described here, is to focus on a single drummer but examine their ``output'' over the course of a series of performances. We have recently been made aware of a suitable archive of isolated hi-hat (and other drum) tracks over the course of a world tour, where songs were played live both with and without a click track (i.e metronome) \citep{Dave}. An analysis of the character of fluctuations for a given drummer performing a specific song over the course of a tour will provide key insights into the extent of the variability of the Hurst exponent(s).

\section*{Data Availability Statement}
The data that support the findings of this study are openly available in the University of Nottingham data repository at https://rdmc.nottingham.ac.uk/handle/internal/348. The associated DOI is http://dx.doi.org/10.17639/nott.344.

\section*{Acknowledgements}

There were a large number of participants in this study to whom we are extremely grateful. We first thank all of those who contributed and/or contacted us via Sixty Symbols (and the associated drumsciproject@gmail.com account), in alphabetical order:

Tommi Aalto, Aaron Addorisio, Mike Arst, Dan Barfoot, Oliver Bleton, Berkay B\"or, Maggie Bishop, Isaac Brightman, Paulo Bucho, Adrian Carrillo, Charlie Carter, Oli	Chandler, Diego Chagoya, Alex Clayden-Spence, David Cornejo, David Corriveau
St-Louis, Joshua Cross, Toby de Mendon¸cam, Guilherme de Mesquita, Pablo de Nadai, Harke de Vlas, Mitch Deighton, Keith E Dennis, Neil Dickson, Davide Donatellis, Phil Douglas, Bryan Edgecombe, Russ Etheridge, Ricardo Esteban Arcila F, Bryan Fink, Matt Forfer, Darren Foulds, Joshua Friedmann, Ernie G, Rich Gibbs, Andrew Goodsell, Eric Gulseth, Jesse Harris, Patrick Stephen Hatch, Anuar Hern\'andez, Tadeus Hogenels, Senanu Koku, Jo\~ao Kopps, Philip Kraus, Alexander Kuzma, John Lamb, Penny Larson,	Dawid Laszuk, Greg Lawrence, Thierry Legroux, Jordan Lerner, Samiel Levin, Cameron Lynn, Jim MacPherson-Amador, Gregory Malanga, Geoffrey Martin, Adan Martinez (audio engineer: Alex Cabral), Kyle Mcaslan, Callum Morrison, Jaakko Niemi, Jesse Nikolic, Max Nippard, Tyler Paul, Matthew Pediaditis, Oliver Petri, Mark Poulter, Nicholas Prabaharan, Zack Prim, Hendrik Richter, Frederik Rosenkjær, Miguel Salvatierra Reyes, Zachary Shumar, Sina, Liam Singleton, Kajetan Stanski, Phil Stockton, Tristan Strahler, Jason Stennett, Saku Taittonen, Shawhin Talebi, Colin Taylor, ``The Demon Lord'', Mass\'e Thomas, Ryan Trias, Stacy Vasquez, Peter Vermette,	Viktor Volari\'c, Max Welsman, Matt Wiebke, Mitch Wilkinson, Alex Woodburn, Piotr Zalubski, and Erik Ziak.

The bulk of the analysis, however, was only possible through the assistance of the following drummers who agreed to having their hi-hat tracks recorded: Jonathan Gosling, Andy McKeown, and Emma Stafford (University of Nottingham); Molly Parker, Cameron, Sam Croft, Harvey Neizer-arku, Lili Popper, George Hadden, Janice Chan, and Bua Nateejarurat (Nottingham-Trent University); Aaron Smith, Steve	Barwell, Ben Miles, Ben Beilby, Aaron Dolton, Ashley Matthews, Dave Hickman, and
Paul Hodson (Staffordshire University); Karen Gourlay, Connor Small, Nathan Birch, Ellie M, Herbie Allen, Tom Gibbins, Matt Martin, Louis Tissiman, Tom Bortlik, and Eliza Wheatley (Saturday Music School and Leeds College of Music).

Complementing the drummers we were fortunate to have a group of secondary school (Year 12/Year 13) students without drumming experience who were willing to play a hi-hat along to a metronome running at 88 bpm. Their efforts are thoroughly
appreciated as they provided an important “control” for our DFA analysis.

Richard Lee generously provided Aerodrums versions of both \textit{Tom Sawyer} and \textit{I Keep Forgettin’} and also gave the manuscript a critical reading, for which we are very
grateful. We also gratefully acknowledge the advice and help of both Esa R\"as\"anen and Hennig Holger throughout the preparation of this paper (including generously sharing
data and code). Finally, we would like to thank both Brady Haran and Sean Riley for bringing the project to a wider audience via the Sixty Symbols channel and for their  important input to framing the study for a non-academic audience.

\newpage

\bibliographystyle{apacite}
\bibliography{RushingOrDragging}

\end{document}